\documentstyle[12pt]{article}
\date{}
\begin{document}

{\large
{\bf
\begin{center}
Reductions of the Volterra and Toda chains
\end{center}
}
}

{\large
\begin{center}
Andrei K. Svinin\footnote{E-mail:svinin@icc.ru}\\
Institute for System Dynamics and Control Theory\\
Siberian Branch of Russian Academy of Sciences\\
P.O. Box 1233, 664033 Irkutsk, Russia
\end{center}
}

\begin{abstract}
The Volterra and Toda chains equations are considered. 
A class of special reductions for these equations are derived.
\end{abstract}

\newpage

The Volterra and Toda chains are known to have numerous applications.
In particular, they appear in connection with continuum two-boson KP 
hierarchies \cite{aratyn} and their discrete symmetries (or auto-B\"acklund
transformations). It is worthing also to mention the works \cite{levi},
\cite{leznov} where integrable chains serve as discrete symmetries
of differential evolution equations.

In this note we establish finite-dimensional systems arising in process
of special reductions of Volterra and Toda lattices. Translations on the
lattices, as in the case of two-boson KP hierarchies, yield groups of discrete
symmetries for these finite-dimensional systems.  

Consider the Volterra lattice in the form 
\begin{equation}
\dot{r}(i) = r(i)\left\{r(i-1) - r(i+1)\right\},\;\; i\in{\bf Z}.
\label{Volterra}
\end{equation}
For any fixed $n\in{\bf N}$, impose a constraint in the form of the
following algebraic equations:
\begin{equation}
r(i) + ... + r(i+n-1) = r(i-1)r(i)...r(i+n),\;\; i\in{\bf Z}.
\label{constraint1}
\end{equation}
We are going to show that infinite collection of equations (\ref{Volterra})
restricted by relations (\ref{constraint1}) are equivalent to 
finite-dimensional systems complemented by auto-B\"acklund transformations
(ABT).

Fix any $i = i_0\in{\bf Z}$ and define finite collection of functions
${\bf y} = (y_1(t),..., y_{n+1}(t))$ by
$$
y_1 = r(i_0),\; y_2 = r(i_0+1),..., y_{n+1} = r(i_0+n).
$$
Taking into account (\ref{constraint1}), we easy extract finite-dmensional
systems
\begin{equation}
\begin{array}{l}
\displaystyle
\dot{y}_1 = \frac{y_1 + ... + y_n}{y_2...y_{n+1}} - y_1y_2,\\[0.3cm]
\displaystyle
\dot{y}_k = y_k(y_{k-1} - y_{k+1}),\;\; k = 2,..., n,\\[0.3cm]
\displaystyle
\dot{y}_{n+1} = y_ny_{n+1} - \frac{y_2 + ... + y_{n+1}}{y_1...y_{n}}.
\end{array}
\label{finite1}
\end{equation}
With any solution of Volterra chain constrained by (\ref{constraint1}),
for different $i_0$, we should have respectively different solutions
of the system (\ref{finite1}). If the constraint (\ref{constraint1})  
is consistent with Volterra chain (\ref{Volterra}) then the shift
$i_0 \rightarrow i_0 + 1$ must yield invertible ABT for the system
(\ref{finite1}). Define ``new" functions
$$
\tilde{y}_1 = r(i_0+1),\; \tilde{y}_2 = r(i_0+2),..., 
\tilde{y}_{n+1} = r(i_0+n+1).
$$
From (\ref{constraint1}), we easy obtain
\begin{equation}
\tilde{y}_1 = y_2,..., \tilde{y}_n = y_{n+1},\;\;
\tilde{y}_{n+1} = \frac{y_2 + ... + y_{n+1}}{y_1y_2...y_{n+1}}.
\label{ABT1}
\end{equation}
By straightforward but tedious calculations, one can verify that
if $(y_1(t),..., y_{n+1}(t))$ is some solution of (\ref{finite1})
then, by virtue (\ref{ABT1}), $(\tilde{y}_1(t),..., \tilde{y}_{n+1}(t))$
also will be solution of (\ref{finite1}). So we can conclude that
the mapping (\ref{ABT1}), for the system (\ref{finite1}), serve as
discrete symmetry. It is easy to deduce inverse to (\ref{ABT1}). We
have
$$
y_{1} = \frac{\tilde{y}_1 + ... + \tilde{y}_{n}}{\tilde{y}_1
\tilde{y}_2...\tilde{y}_{n+1}},\;
y_2 = \tilde{y}_1,..., y_{n+1} = \tilde{y}_{n}.
$$

One defines
$$
a_0(i) = - r(2i-1) - r(2i),\;\; a_1(i) = r(2i-2)r(2i-1).
$$
These relations are known to realize the map from the solution
space of (\ref{Volterra}) to that of Toda chain equations \cite{Toda},
\cite{wadati}
\begin{equation}
\dot{a}_0(i) = a_1(i+1) - a_1(i),\;\;  
\dot{a}_0(i) = a_1(i)\left\{a_0(i) - a_0(i-1)\right\},\; i\in{\bf Z}.
\label{TL1}
\end{equation}
Now observe that, for even $n$, replacing in (\ref{constraint1})
$i \rightarrow 2i - 1$, these equations can be rewritten in the form
\begin{equation}
- a_0(i) - ... - a_0(i+m-1) = a_1(i)a_1(i+1)...a_1(i+m),\; i\in{\bf Z}
\label{constraint2}
\end{equation}
where $m = n/2$.

It is naturally to suppose that constraint (\ref{constraint2}) also will lead
to finite-dimensional systems. To write Toda lattice (\ref{TL1}) in
more familiar form, one introduces the functions $\{u_i(t),\; i\in{\bf Z}\}$
by
$$
a_0(i) = - \dot{u}_i,\;\;a_1(i) = e^{u_{i-1} - u_i}.
$$    
After that we have
\begin{equation}
\ddot{u}_i = e^{u_{i-1} - u_i} - e^{u_i - u_{i+1}},\;\; i\in{\bf Z}.
\label{TL}
\end{equation}
The constraint (\ref{constraint2}) becomes
\begin{equation}
\dot{u}_i + ... + \dot{u}_{i+m-1} = e^{u_{i-1} - u_{i+m}},\;\; i\in{\bf Z}.
\label{constraint}
\end{equation}
Observe that, in terms of $\tau$-function defined through
$$
e^{u_i} = \frac{\tau_i}{\tau_{i+1}},
$$
the equation (\ref{constraint}) takes simple form. Namely, we have bilinear
equation
$$
D_t\tau_i\cdot\tau_{i+m} = \tau_{i-1}\tau_{i+m+1}.
$$
Recall that Hirota's $D$-operator is defined by
$$
D_t^lf\cdot g = (\partial_{t_1} - \partial_{t_2})^lf(t_1)g(t_2)|_{t_1=t_2=t}.
$$ 

By analogy with Volterra lattice situation, introduce finite collection
of functions ${\bf q} = (q_1(t),..., q_{m+1}(t))$ identifying
$$
q_1 = u_{i_0},\; q_2 = u_{i_0+1},..., q_{m+1} = u_{i_0+m}.
$$
Taking into account the relation (\ref{constraint}), we easy obtain
finite-dimensional systems
\begin{equation}
\begin{array}{l}
\ddot{q}_1 = (\dot{q}_1 + ... + \dot{q}_{m})e^{q_{m+1}-q_1} - e^{q_1-q_2},\\[0.3cm]
\ddot{q}_k = e^{q_{k-1}-q_k} - e^{q_k-q_{k+1}},\;\; k = 2,..., m,\\[0.3cm]
\ddot{q}_{m+1} = e^{q_{m}-q_{m+1}} - (\dot{q}_2 + ... + \dot{q}_{m+1})e^{q_{m+1}-q_1}.
\end{array}
\label{finite}
\end{equation}
Define
$$
\tilde{q}_1 = u_{i_0+1},\; \tilde{q}_2 = u_{i_0+2},..., \tilde{q}_{m+1} = u_{i_0+m+1}.
$$
From (\ref{constraint}), we easy deduce the relations
\begin{equation}
\tilde{q}_1 = q_2,..., \tilde{q}_{m} = q_{m+1},\; \tilde{q}_{m+1} = q_1 - 
\ln\left[\dot{q}_2 + ... + \dot{q}_{m+1}\right].
\label{ABT}
\end{equation}
By straightforward calculations, one can check that transformation
(\ref{ABT}), for any $m\in{\bf N}$, realize ABT for corresponding
system. It is easy to write down inverse to (\ref{ABT}). We have
$$
q_1 = \tilde{q}_{m+1} + \ln\left[\dot{\tilde{q}}_1 + ... + \dot{\tilde{q}}_{m}
\right],\;\; q_2 = \tilde{q}_1,..., q_{m+1} = \tilde{q}_{m}. 
$$
We summarize above in the following statement.

{\bf Proposition 1.} The equations of Volterra and Toda lattices restricted,
respectively, by relations (\ref{constraint1}) and (\ref{constraint})
are equivalent to the system (\ref{finite1}) and (\ref{finite}) complemented 
by ABT (\ref{ABT1}) and (\ref{ABT}).  

It is easy to get the relations connecting the solution spaces of the systems
(\ref{finite1}) and (\ref{finite}), in the case when $n=2m$. We have
$$
y_{2k-1} + y_{2k} = \dot{q}_k,\;\; k = 1,..., m,
$$
$$
y_{2m+1} + \frac{y_2 + ... + y_{2m+1}}{y_1y_2...y_{2m+1}} = \dot{q}_{m+1},
$$
$$
y_{2k}y_{2k+1} = e^{q_k - q_{k+1}},\;\; k = 1,..., m.
$$

Observe that any system (\ref{finite}) can be cast into Lagrangian formalism.

{\bf Proposition 2.} The systems (\ref{finite}) admit Lagrangian formulation
with appropriate Lagrangian
$$
{\cal L} = \sum_{i<j}\dot{q}_i\dot{q}_j 
- \sum_{i=1}^{m}e^{q_i - q_{i+1}} -
(\frac{1}{2}\dot{q}_1 + \sum_{j=2}^{m}\dot{q}_j + \frac{1}{2}\dot{q}_{m+1})
e^{q_{m+1} - q_1}.
$$

{\it Proof.} Calculate Euler derivatives of ${\cal L}$
$$
E_{q_i}{\cal L} = \frac{\partial{\cal L}}{\partial q_i} - 
\frac{d}{dt}\left(
\frac{\partial{\cal L}}{\partial \dot{q}_i}
\right).
$$
We have 
$$
E_{q_1}{\cal L} = - \sum_{j=2}^{m+1}\ddot{q}_j - e^{q_1 - q_2} +
\left(\sum_{j=2}^{m+1}\dot{q}_j\right)e^{q_{m+1}-q_1},
$$
$$
E_{q_k}{\cal L} = - \sum_{j\neq k}\ddot{q}_j - e^{q_k - q_{k+1}} +
e^{q_{k-1} - q_k} + 
$$
$$
+ (\dot{q}_{m+1} - \dot{q}_1)e^{q_{m+1}-q_1},\;\; k = 2,..., m,
$$
$$
E_{q_{m+1}}{\cal L} = - \sum_{j=1}^m\ddot{q}_j + e^{q_m - q_{m+1}} -
\left(\sum_{j=1}^{m}\dot{q}_j\right)e^{q_{m+1}-q_1},
$$

Equating $E_{q_i}{\cal L}$ to zero results in the system of the form
\begin{equation}
\sum_{j=1}^{m+1} A_{ij}\ddot{q}_j = f_i(q_k, \dot{q}_k)
\label{A}
\end{equation}
where the matrix $A = (A_{ij})$ consists of zeros in main diagonal
and units elsewhere. It is easy to see that $A$ is nondegenerate.
Moreover it can be checked that $A^{-1}$ consists of $(2-m-1)/m$'s
in main diagonal and $1/m$'s elsewhere. One can verify that substituting 
right-hand sides of the system (\ref{finite}) in (\ref{A}) gives identity.
So we can conclude that (\ref{finite}) and (\ref{A}) are equivalent. $\Box$   

It is simple exercise, using Legendre transformation, to cast any
system (\ref{finite}) in Hamiltonian setting. In the simplest case $m=1$, 
(\ref{finite}) becomes
\begin{equation}
\ddot{q}_1 = \dot{q}_1e^{q_2-q_1} - e^{q_1-q_2},\;\;
\ddot{q}_2 = e^{q_1-q_2} - \dot{q}_2e^{q_2-q_1}.
\label{system}
\end{equation}
Its ABT reads
\begin{equation}
\tilde{q}_1 = q_2,\;\; \tilde{q}_2 = q_1 - \ln(\dot{q}_2).
\label{ABT3}
\end{equation}

Note two first integrals for the system (\ref{system}). They are
$$
P = -\dot{q}_1 - \dot{q}_2 - e^{q_2 - q_1},\;\;
E = -\dot{q}_1\dot{q}_2 + e^{q_1 - q_2}.
$$
It can be checked that $P$ and $E$ are invariant with respect to 
ABT (\ref{ABT3}).

Lagrangian in this case looks as
$$
{\cal L} = -\dot{q}_1\dot{q}_2 - e^{q_1 - q_2} - 
\frac{1}{2}(\dot{q}_1 + \dot{q}_2)e^{q_2 - q_1}.
$$
We easy obtain
$$
p_1 = \frac{\partial{\cal L}}{\partial\dot{q}_1} = 
- \dot{q}_2 - \frac{1}{2}e^{q_2 - q_1},\;\;
p_2 = \frac{\partial{\cal L}}{\partial\dot{q}_2} = 
- \dot{q}_1 - \frac{1}{2}e^{q_2 - q_1},
$$
$$
{\cal H} = 
\dot{q}_1\frac{\partial{\cal L}}{\partial\dot{q}_1} +
\dot{q}_2\frac{\partial{\cal L}}{\partial\dot{q}_2} -
{\cal L} = - (p_1 + \frac{1}{2}e^{q_2 - q_1})(p_2 + \frac{1}{2}e^{q_2 - q_1}) +
e^{q_1 - q_2}
$$
It is easy to check that the functions ${\cal H}$ and $P = p_1 + p_2$ are 
in involution with respect to standard Poisson bracket. So we can conclude
that the system (\ref{system}) is integrable due to Liouville's theorem.
It is naturally to suppose that all systems (\ref{finite}) are completely
integrable in the sense of Liouville's theorem \cite{liouville}.

\end{document}